\documentclass[]{aa} 

\usepackage{graphicx}
\usepackage{txfonts}
\begin{document}
\title{Spectrophotometric analysis of the 5200\AA\ region for peculiar and normal stars }

\author{Ch.~Stigler\inst{1}
\and H.M.~Maitzen\inst{1}
\and E.~Paunzen\inst{2}
\and M.~Netopil\inst{1,2}}
\institute{
Institut f{\"u}r Astrophysik der Universit{\"a}t Wien, T{\"u}rkenschanzstr. 17, A-1180 Wien, Austria \\
\email{hans.michael.maitzen@univie.ac.at}
\and Department of Theoretical Physics and Astrophysics, Masaryk University,
Kotl\'a\v{r}sk\'a 2, 611\,37 Brno, Czech Republic \\
\email{epaunzen@physics.muni.cz}}

   \date{} 
  \abstract
   {Many chemically peculiar (CP) stars, especially the magnetic CP2 stars, show a flux depression at 5200\AA. The $\Delta a$ photometric system
   takes advantage of this characteristic to detect these objects in an efficient way. In addition, it is capable of finding metal-weak, emission-type, and
   shell-type objects of the upper main sequence.}
   {To compare available observations and to detect new peculiar objects, we used a spectrophotometric catalogue consisting of 1159 stars. From this catalogue, 
	  we selected 1067 objects to synthesize three different $a$ indices to find the most efficient one for further observations. In addition, we extended the analysis 
    to stars cooler than F5.  }
   {We employed classical $\Delta a$ photometry described by Maitzen, using simulated filter curves, the spectrophotometric $\Delta a$ index by Adelman, and a modified index.}
{Even though the accuracy of the spectrophotometry used for this investigation is significantly lower than the photometric $\Delta a$ measurements, we are able to confirm peculiarity for most of the known CP2 stars above a certain limit of $\Delta a$. We investigated 631 stars hotter than spectral type F5 to find additional that are not yet identified peculiar objects.
We find that for very low mass stars (M0), the $a$ index is independent of the colour (effective temperature).}
  {The $\Delta a$ photometric system is very closely correlated with the effective temperature over a wide range of the main sequence. It is able to detect any kind of peculiarity connected
  to the 5200\AA\ region. Especially for low-mass stars, this opens up a new possibility of detecting peculiar objects in an efficient way.}

\maketitle
%
\section{Introduction}

Chemically peculiar (CP hereafter) stars of the upper main sequence differ in their abundances 
of heavy and rare-earth elements in their photosphere from normal-type objects of the same effective temperature
and luminosity range. \citet{Pres74} divided this group into four subgroups according to their
astrophysical characteristics. Since then several investigations about their local
environment and stellar parameters were performed. \citet{Babc47} discovered a global dipolar magnetic 
field in the star 78 Virginis, which was soon followed by the detection of several other similar stars. 
Therefore the correlation of stellar magnetic field strengths with astrophysical processes such as diffusion and
meridional circulation can be very well studied in this stellar group \citep{Glag13}. Another important often studied 
aspect is the cause of the peculiar (surface) abundances. The observed phenomenon can be explained by either classical diffusion of chemical elements
depending on the balance between gravitational pull and uplift by the radiation field through absorption
in spectral lines \citep{Stif12} or selective accretion from the interstellar medium via the stellar magnetic field 
\citep{Havn71}.
While diffusion seems to be appropriate for both magnetic and
non-magnetic stars to explain spectral peculiarity, it is not yet clear
to which extent the interaction with the interstellar medium via accretion
and transport of angular momentum may modify the effects of diffusion 
and break the stellar rotational velocities during the stars' main sequence life time.

Using spectrophotometry, \citet{Koda69} was the first to notice broad-band flux depressions at 4100, 5200 and 6300\AA\ during his 
investigation of the magnetic CP (CP2) star HD 221568. Photometrically, the main depressions (4100 and 5200\AA) of another 
CP2 star were later also found by \citet{Mait72} when they investigated the spectrum-variable HD 125248. 
This resulted in the development of the $\Delta a$ photometric system \citep{Mait76}, which makes use of the most efficient depression at 5200\AA . The flux depression itself
is a combined contribution of Si, Cr and Fe for strongly overabundant surface metallicities \citep{Khan07}. The upper
main sequence is unique for CP stars. No detailed observational analysis of this region for cooler objects has been
performed so far.

In this paper, we present a study that was conducted to detect CP stars on the basis of spectrophotometric data for which we employed three different 
methods that use the flux depression at 5200\AA. In addition to the original method, the $a'$ system developed by \citet{Adel79} 
and a newly modified method were investigated. Furthermore, we investigated the capability of these systems of finding new, previously
undetected, peculiar stars including metal-weak and emission-type objects of the upper main sequence. For the first time, the
low-mass regime was searched for the behaviour of the $a$ indices in correlation with the effective temperature and luminosity.

\begin{figure}
\begin{center}
\includegraphics[width=85mm]{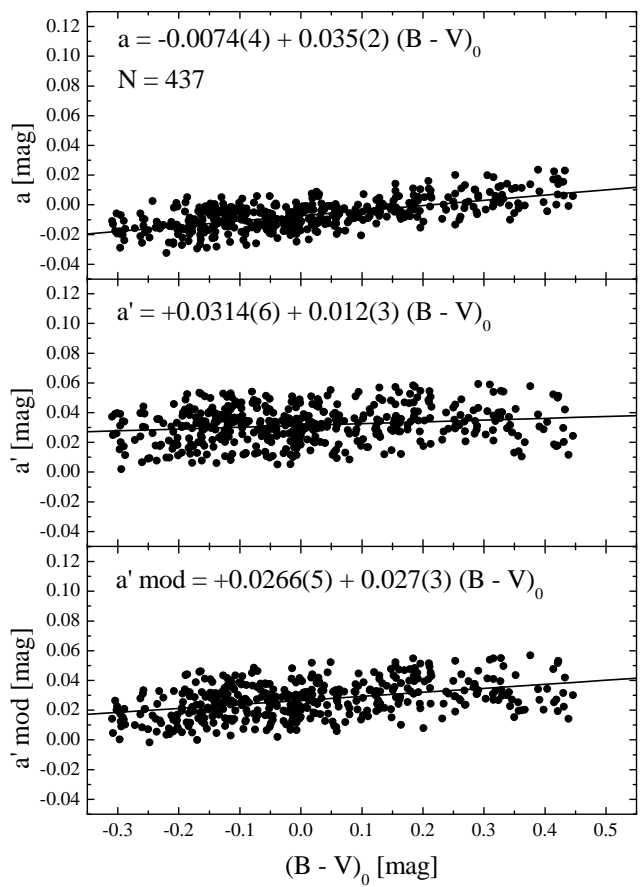}
\caption{Regression of 437 normal stars with reference to the different $a$ indices.}
\label{regression}
\end{center}
\end{figure}

\begin{figure}
\begin{center}
\includegraphics[width=9cm]{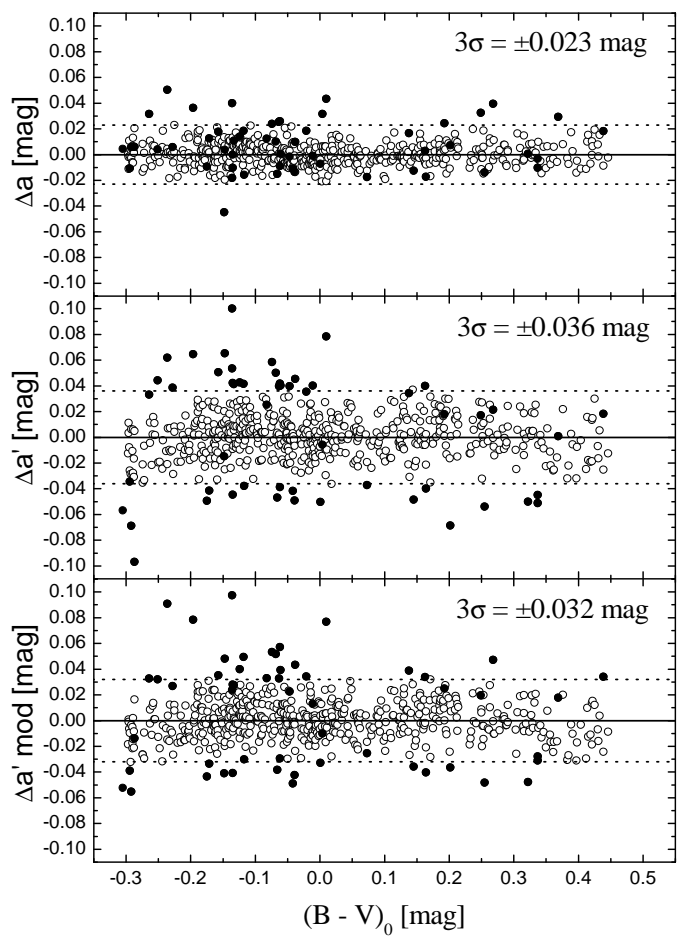}
\caption{Peculiarity indices of 631 stars in the spectral-class range from O6 to F5. The filled circles denote stars that
are detected as peculiar within at least one index.}
\label{3sigma}
\end{center}        
\end{figure}
   
\section{Data and reduction}

Our source of spectrophotometric data is the stellar catalogue by \citet{Khar88}. It contains data for 1159 bright 
stars of different spectral classes in the wavelength range from 3200 to 7600\AA, with a spectral resolution of 50\AA. Each star was observed 
at least three times, on different nights. The root mean square (rms hereafter) relative error of the measurements 
is between 2\% and 4\% per 50\AA\ bandpass. Stars with only two-digit measurements in the wavelength region around 5200\AA\ 
were not considered in the analysis because of the statistical inaccuracy of these measurements. We also excluded early-type
supergiants because of their strong photometric and spectroscopic variability. The remaining data of 1067 stars
were used to synthesize different peculiarity indices. The methods used to detect CP stars are designed to have magnitudes as input. Therefore the 
intensity and flux values given in the catalogue were converted into magnitudes.

We are aware that there are more extensive corresponding catalogues with higher resolutions available in the
literature, for example the one by \citet{Burn85}. We note that this author also used at that time unpublished data by \citet{Khar88}, but
interpolated the data to a resolution of 25\AA. However, the catalogue we used is the basis of a list of spectrophotometric
standards published by \citet{Glus92}. Therefore the quality of the data is beyond any doubt. In addition, the spectral resolution is
similar to that used for the Gaia mission \citep{Jord10}.

Spectrophotometric observations of classical CP stars were published before, for example by \citet{Mait80b} and \citet{Adel89}. However,
all these analyses concentrated either on individual stars or on the comparison with results from synthetic spectra. To our knowledge, no systematic
analysis of spectrophotometric observations with respect to synthesizing $\Delta a$ was conducted before.
   
The goal of this paper is not only to detect CP stars, but also to compare our findings with already existing 
$\Delta a$ photometry and spectrophotometry. For this purpose, we employed three different systems. 

{\it The $\Delta a$ photometric system by \citet{Mait76}:} this system compares the flux in the centre of the depression (filter $g_{\rm 2}$, 
$\lambda_{c}$\,=\,5220\AA) with the adjacent spectral regions (filters $g_{\rm 1}$, 5000\AA\ and $y$, 5500\AA) using a band-width of 130\AA, which represents the 
continuum of the star. With these measurements the index $a$ can be calculated as
\begin{equation}
a = g_{\mathrm{2}} - \frac{g_{\mathrm{1}}+y}{2}.
\end{equation} 
In principle, the positioning of the continuum filters $g_{\mathrm{1}}$ and $y$ relative to $g_{\mathrm{2}}$ minimizes the 
influence of temperature on the peculiarity index. However, due to slight wavelength mismatches of the filters, a second-order dependence 
on temperature is apparent in addition to a general increase of opacity around 5200\AA\ with decreasing temperature.
Therefore, one has to normalize $a$ with the index $a_{\mathrm{0}}$ of a non-peculiar star of the same temperature (normality line), 
to compare different peculiar stars with each other. The photometric peculiarity index is therefore
\begin{equation}
\Delta a = a-a_{\mathrm{0}}.
\end{equation} 
To do so, the corresponding filter curves had to be simulated. 
For simplicity, the three filters were represented by Gaussian curves with a FWHM of 130\AA\ and a transmission
maximum of 100\% each. 
The fact that these simulated filters, in contrast to the original definition of $g_{\mathrm{1}}$, $g_{\mathrm{2}}$ and $y$, 
all have the same transmission curves, probably does not significantly affect $\Delta a$, because its value is always 
the difference between a peculiar star and the corresponding normal star with the same colour index. To convolve the 
Gaussian curves with the measurements of the spectrophotometric catalogue, transmission values of 50\AA\ (bins) of 
the simulated filters, corresponding to the measurements with 50\AA\ spectral resolution, had to be determined. 
This was achieved by segmenting the Gaussian curves into 5\AA\ bins, which were then numerically integrated. 
The percentage of transmission, relative to the maximum, could now be calculated by applying the mean value 
theorem. For the transmission values of the 50\AA\ bins, to be multiplied with the measurements, simply ten of 
the 5\AA\ bin transmission values were summed. This approach enabled us to adapt the central wavelengths of 
the filters in 5\AA\ steps to figure out the optimal filter positions. The central wavelengths of the filters 
were optimised with respect to a low variance of $a$ values for normal stars and secondly the highest possible 
$a$ values of CP2 stars. The best compromise of filter positions found for the spectrophotometric data is 
5020\AA\ for $g_{\mathrm{1}}$, 5215\AA\ for $g_{\mathrm{2}}$, and 5470\AA\ for $y$. These filter positions 
are very close to those of the revised $\Delta a$ system by \citet{Mait80a}. We calculated the values of the simulated filters
by multiplying the transmission values of the 50\AA\ bins with the corresponding measurements of the catalogue 
and adding the results.

\begin{figure}[h]
\begin{center}
\includegraphics[width=9cm]{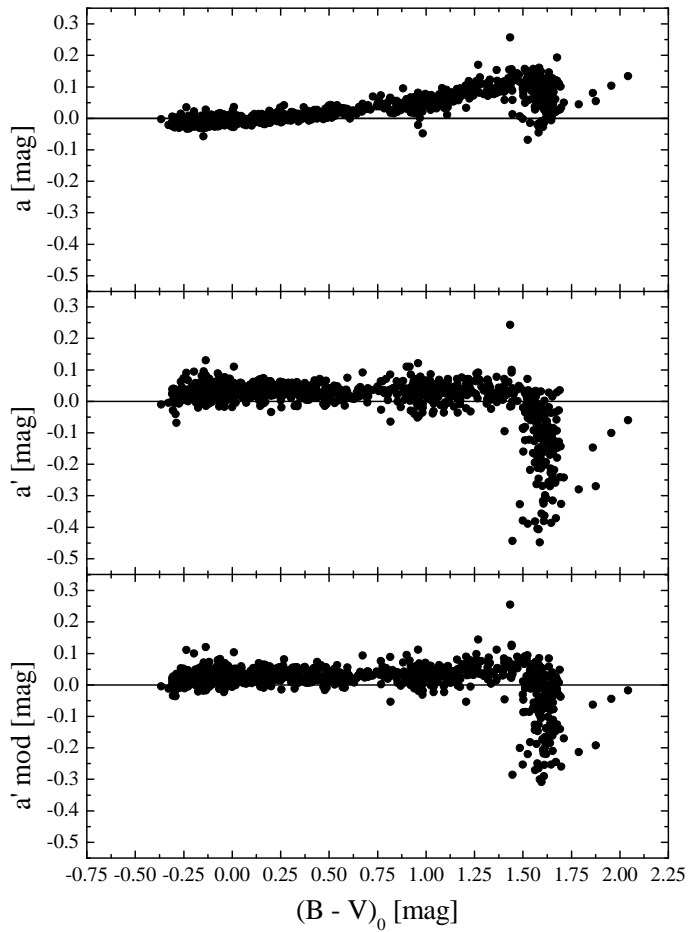}
\caption{Different $a$ indices of 1067 stars from the catalogue by \citet{Khar88}
plotted versus $(B-V)_0$.}
\label{a_bv}
\end{center}
\end{figure}

{\it The $\Delta a'$ index by \citet{Adel79}}: it basically 
works in the same way as the index by Maitzen, but was especially designed for his spectrophotometric data. It is defined as
\begin{equation}
a' = m_{\mathrm{5264}} - [m_{\mathrm{4785}} + 0.453(m_{\mathrm{5840}} - m_{\mathrm{4785}})].
\end{equation}
The measurements at 4785 and 5840\AA, which were both made at locations significantly bluer than $g_{\mathrm{1}}$ and redder than $y$, 
represent a wavelength base for the continuum that is larger by a factor of two than in $\Delta a$ photometry. 
The depression itself is represented by the 5264\AA\ value. Like the $a$ index by Maitzen, the $a'$ index is slightly dependent 
on the colour of the star and therefore has to be normalized by the index of non-peculiar stars with the same colour.

{\it The $\Delta a'mod$ index:} Very early in the process of evaluating the spectrophotometric data the central wavelengths of the 
spectrophotometric data did not fit those of the $\Delta a'$ index very well. To calculate the 
$a'$ index, the measurements closest to the defined wavelengths were used. This resulted in inaccurate results, because
the central measurement was shifted from the deepest part of the depression to the 
red side of the spectra and in some stars the $H\beta$ line influenced the continuum value on the 
blue side. To address this problem two values next to each other were linearly interpolated to obtain central 
wavelengths closer to the predefined wavelengths of the $\Delta a'$ index, as well as to improve the error 
by a factor of $\sqrt{2}$. This new index is referred to as $\Delta a'mod$ and is defined as
\begin{equation}
a'mod = m_{\mathrm{5250}} - [m_{\mathrm{4750}} + 0.453(m_{\mathrm{5850}} - m_{\mathrm{4750}})].
\end{equation}

The calculation of the indices $a'$ and the $a'mod$ was straightforward. They were obtained by simply 
applying the values of the corresponding measurements to the formula. 

As next step, the indices had to be corrected with the $a$, $a'$, and $a'mod$ obtained from non-peculiar 
stars of the same temperature. For this purpose, all stars hotter than F5 were selected to 
calculate $a_{\mathrm{0}}$, $a'_{\mathrm{0}}$ and $a'_{\mathrm{0}}mod$. 

According to \citet{Mait76}, the indices are expected to be very well correlated with the colour index $(B-V)_0$. 
To deredden the programme stars, we made use of photometric data in the Johnson, Geneva, and Str{\"o}mgren systems, compiled from 
the General Catalogue of Photometric Data (GCPD)
\citep{Merm97}.
For the first two systems, the well-known calibrations based on the X/Y parameters \citep{Cram99}
and Q-index \citep{Guti75}, respectively, which are applicable for O/B type
stars were applied. Objects with available Str{\"o}mgren data were treated with the routines by \citet{na93}, which allow
dereddening of cooler-type stars. For about 60\,\% of the targets we were able to determine reddening values, transformed to $E(B-V)$
with the ratios summarised by \citet{net08}. If several estimates for a particular object were available, a mean value was calculated.
Since all objects are rather bright and therefore close-by, a strong reddening especially for cool-type stars is hardly expected.
We therefore adopted non-reddening for these when no Str{\"o}mgren photometry was available, which is justified by the obtained overall mean $E(B-V)=0.02(6)$\,mag.

To find the normality lines, we performed an iterative linear regression of $a$, $a'$ and $a'mod$ versus
$(B-V)_0$. In each iterative step, we discarded the outliers that lay more than 5$\sigma$ from the
normality line. After three iterations, the errors did not significantly decrease any more. In total, we
defined the normality lines from 437 stars and obtained a 3$\sigma$ of 23, 36, and 32\,mmag, respectively.
Figure \ref{regression} shows the regression of the peculiarity indices versus $(B-V)_0$.

The three $\Delta a$ indices were calculated by subtracting the corresponding
$a_{\mathrm{0}}$, $a'_{\mathrm{0}}$ and $a'_{\mathrm{0}}mod$ from the $a$, $a'$ and $a'mod$ values for the 
same $(B-V)_0$. 
Figure \ref{regression} shows the peculiarity indices
versus $(B-V)_0$. All stars above the threshold of 3$\sigma$ are expected to be chemically peculiar, whereas those below are
emission-type and/or metal-weak objects. 
The region of $(B-V)_0$ in which CP2 stars are expected (hotter than spectral type F5) contains 631 stars (Fig. \ref{3sigma}).

At $(B-V)_0$ of about 1.5 mag, corresponding to a spectral type M0, the 
indices $a$, $a'$ and $a'mod$ lose their correlation with $(B-V)_0$ although they are nearly linearly correlated 
at bluer colour indices.
This may indicate a dependency of $a$, $a'$ and $a'mod$ on the luminosity class of the stars (see Fig. \ref{a_bv}).
   
\section{Results}

The findings of our investigation are summarized in Tables \ref{deltaa_values_cps} and \ref{deltaa_values_other}. 
The first table lists all well-known CP stars taken from the catalogue by \citet{rens09}. The second table
includes apparently normal-type objects (according to spectral classification) hotter than spectral type F5 that have $\Delta a$ values both higher and lower than 
the 3$\sigma$ thresholds. The tables contain the HD number, spectral type, the $\Delta a$ values of each applied method,
and the CP class (Table \ref{deltaa_values_cps} only) of the stars. For comparison with existing peculiarity measurements, they also 
include published $\Delta a$ values \citep{Paun05} and $\Delta (V1-G)$ values. 
The $\Delta (V1-G)$ index is a measurement for peculiarity derived from the Geneva photometric system. \citet{Hauc74} 
was the first to propose an index as peculiarity parameter:

\begin{equation}
\Delta (V1-G) = (V1-G)-0.289(B2-G)+0.302.
\end{equation}

On average, normal stars have $\Delta (V1-G)$ values of $-$5 mmag, therefore we added +5 mmag to the calculated 
values. The photometric data of the Geneva system were collected by using the GCPD. 

\begin{figure}
\begin{center}
\includegraphics[width=80mm]{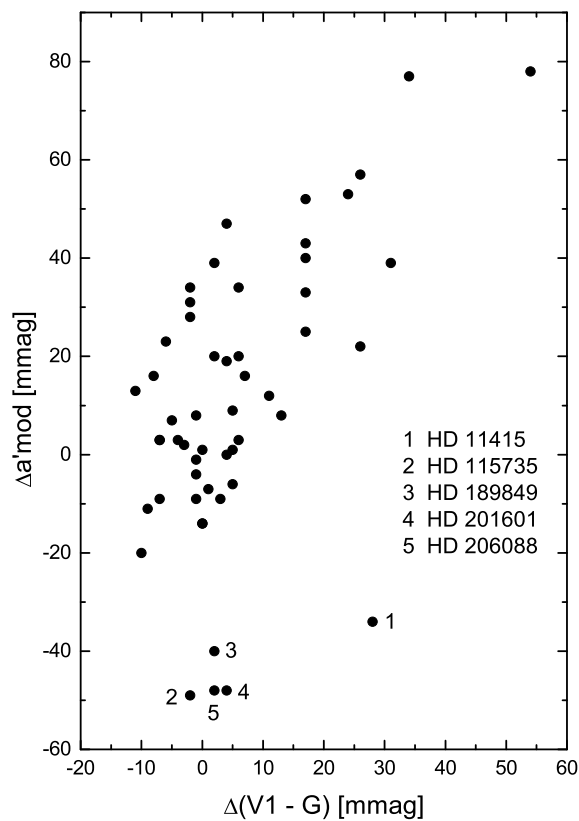}
\caption{Observed $\Delta (V1-G)$ versus synthetic $\Delta a' mod$ values ($\Delta a$ and $\Delta a'$ behave very similar)
for the CP star sample. The outliers are discussed in the text.}
\label{synth_versus_obs}
\end{center}
\end{figure}

\begin{figure}
\begin{center}
\includegraphics[width=80mm]{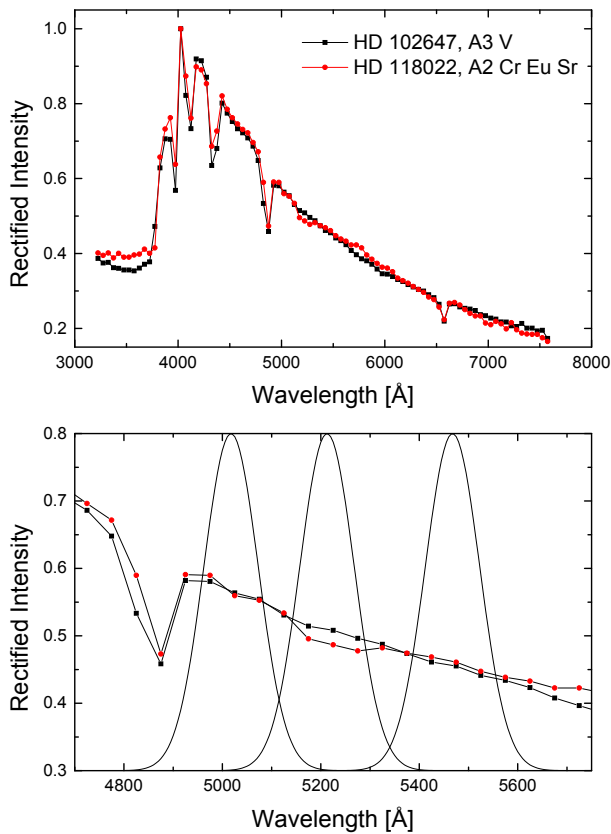}
\caption{Spectrophotometric data of HD 102647 (A3 V) and HD 118022 (A2 Cr Eu Sr) for
the whole spectral range (upper panel) and the region where the $\Delta a$ system is situated (lower panel). 
The UV excess and the 5200\AA\ depression, both typical for CP stars, are clearly visible for
HD 118022.}
\label{filtercurves}
\end{center}
\end{figure}

\subsection{Known peculiar objects} \label{cps_sub}

First of all, we analysed the 55 CP stars listed in Table \ref{deltaa_values_cps} that are included in the catalogue
by \citet{rens09}. Altogether, 19 stars were detected beyond the 3$\sigma$ threshold by any of the systems.
The modified system of Adelman yields 18 stars above 3$\sigma$ for this sample. This is the highest ratio 
of spectrophotometric to otherwise identified peculiar stars of the three applied detection methods.

In Fig. \ref{synth_versus_obs} we present the observed $\Delta (V1-G)$ versus the synthetic $\Delta a' mod$ values
for our sample. The observed $\Delta a$ values were not used because there are too few available. There is a clear
correlation of the observed and synthesized values. The five outliers (HD 11415, HD 115735, HD 189849, HD 201601, and HD 206088) 
are described below.

The CP1 (Am) stars have, with some rare exceptions, observed $\Delta a$ values well below +10\,mmag and are normally inconspicuous in the $\Delta (V1-G)$
index. However, we detected four stars, three above and one below, the 3$\sigma$ threshold; these are
\begin{itemize}
\item {\it HD 29479}: a moderate overabundance of Fe-peak elements and [Ba/H]\,=\,+1.77\,dex compared with that of the Sun
was reported by \citet{Ilie06}
\item {\it HD 76756}: $\alpha$ Cancri is one of the prototype Am stars, with no other conspicuous features 
\item {\it HD 173648}: this is a hot Am star with overabundances of most Fe-peak elements, and considerable 
overabundances of Sr, Y, Zr, and Ba as well as some rare earths \citep{Adel99}
\item {\it HD 189849}: a weak magnetic field of about 250\,G was detected on a 20$\sigma$ level \citep{Bych09}, so it
might be misidentified.
\end{itemize}
It seems that some very peculiar CP1 stars are detectable via spectrophotometric observations and our peculiar
indices.

The magnetic CP2 stars are mainly detectable via $\Delta a$ photometry, which is reflected in our results.
Figure \ref{filtercurves} shows the spectrophotometric data of standard star HD 102647 (A3 V) and HD 118022 (A2 Cr Eu Sr) for
the whole spectral range (upper panel) and the region where the $\Delta a$ system is situated (lower panel). 
The UV excess \citep{Soko06} and the 5200\AA\ depression, both typical for CP stars, are clearly visible for HD 118022.
All strong positive photometric detections are reproduced by our synthetic investigation. There are two objects, that
have significant negative synthetic $\Delta a$ values, but have statistically insignificant observational values:
\begin{itemize}
\item {\it HD 201601}: $\gamma$ Equulei is a well-studied rapidly oscillating Ap star with a strong magnetic field 
and a very peculiar abundance pattern \citep{Perr11}.
\item {\it HD 206088}: a very peculiar object, in a transition between CP1 and CP2 shows strong variability in the
infrared \citep{Cata98}.
\end{itemize}
The spectrophotometry of both objects might be severely influenced by the unusual elemental peculiarity and the
variability.

Among the apparent non-magnetic CP3 (HgMn) stars, no detection in any system was found. This is perfectly 
in line with the results published by \citet{Paun05}.

Our sample includes only three well-known CP4 stars of which two (HD 11415 and HD 115735) show significant negative 
synthetic $\Delta a'$	and $\Delta a' mod$ values. Both objects are known to have strong emission lines according to \citet{Koho99} which explains our result.

\begin{table*}
\caption{Synthetic ($\Delta a$, $\Delta a'$, and $\Delta a' mod$) and observed ($\Delta a\,obs$ and $\Delta (V1-G)$)
peculiarity indices in mmags for well-established CP (flag ``$\ast$'') stars taken from \citet{rens09}. The synthetic 
photometric values for detected objects are given in boldface italics.}
\label{deltaa_values_cps}
\begin{center}
\begin{tabular}{lccccccc}
\hline
\hline 
\\
HD	&	SpType	&	$\Delta a$	&	$\Delta a'$	&	$\Delta a' mod$	&	$\Delta a\,obs$ &	$\Delta (V1-G)$	&	CP group \\
\hline
6116	&	A3$-$A9	&		+8	&	+20	&	+16	&	$-$	&	$-$8	&	1\\
20320	&	A2$-$A9	&	$-$6	&	$-$13	&	$-$14	&	$-$3	&	0	&	1\\
27045	&	A3$-$F3	&	$-$3	&	+1	&	$-$4	&	$-$	&	$-$1	&	1\\
27962	&	A1$-$A4	&	+13	&	+26	&	+31	&	$-$	&	$-$2	&	1\\
28355	&	A5$-$F1	&	$-$1	&	+2	&	+2	&	$-$	&	$-$3	&	1\\
29140	&	A3$-$A7	&	+12	&	+25	&	+23	&	$-$	&	$-$6	&	1\\
29479	&	A3$-$A9	&	+6	&	\textit{\textbf{+37}}	&	+28	&	$-$	&	$-$2	&	1\\
40536	&	A4$-$F1	&	$-$3	&	+18	&	+16	&	+7	&	+7	&	1\\
41357	&	A4$-$F2	&	$-$4	&	+6	&	$-$1	&	$-$	&	$-$1	&	1\\
76756	&	A3$-$F1	&	+17	&	+34	&	\textit{\textbf{+39}}	&	$-$	&	+2	&	1\\
116657	&	A1$-$A7	&	$-$6	&	+19	&	+25	&	+10	&	$-$	&	1\\
125337	&	A2$-$A7	&	$-$11	&	$-$8	&	$-$9	&	$-$	&	$-$7	&	1\\
141795	&	A2$-$A8	&	$-$1	&	$-$6	&	+3	&	$-$	&	$-$7	&	1\\
173648	&	A4$-$F1	&	+3	&	\textit{\textbf{+40}}	&	\textit{\textbf{+34}}	& $-$ & $-$2 & 1 \\
173654	&	A2$-$A7	&	+3	&	+17	&	+13	& $-$ & $-$11 & 1 \\
189849	&	A5$-$A9	&	$-$17	&	\textit{\textbf{$-$40}}	&	\textit{\textbf{$-$40}}	&	$-$	&	+2	&	1\\
197461	&	A7$-$F0 dD	&	$-$6	&	$-$10	&	$-$7	&	$-$	&	+1	&	1\\
198743	&	A3$-$F3	&	+8	&	+5	&	$-$9	&	$-$	&	$-$1	&	1\\
207098	&	A5$-$F4 dD	&	$-$4	&	+7	&	+1	&	$-$	&	+5	&	1\\
223461	&	A3$-$F0	&	+3	&	+7	&	+8	&	$-$	&	$-$1	&	1\\
\hline
10221	&	A0 Si Sr Cr	&	+12	&	+25	&	\textit{\textbf{+33}}	& $-$ & +17 & 2 \\
11502/3	&	 A1 Si Cr Sr	& +10	&	\textit{\textbf{+45}}	&	\textit{\textbf{+43}}	&	+39	&	+17	&	2\\
15089	&	A4 Sr	&	$-$1	&	$-$4	&	0	&	$-$	&	+4	&	2\\
19832	&	B8 Si	&	+14	&	\textit{\textbf{+43}}	&	\textit{\textbf{+40}}	&	+10	&	+17	&	2\\
26571	&	B8 Si	&	+2	&	$-$25	&	$-$20	&	+14	&	$-$10	&	2\\
32549	&	B9 Si Cr	&	+10	&	\textit{\textbf{+50}}	&	\textit{\textbf{+52}}	&	+25	&	+17	&	2\\
32650	&	B9 Si	&	+8	&	+12	&	+12	&	$-$	&	+11	&	2\\
34452	&	B9 Si	&	\textit{\textbf{+36}}	&	\textit{\textbf{+65}}	&	\textit{\textbf{+78}}	&	$-$	&	+54	&	2\\
40312	&	A0 Si	& +7	&	+30	&	+25	&	$-$	&	+17	&	2\\
68351	&	A0 Si Cr	&	\textit{\textbf{+24}}	&	\textit{\textbf{+58}}	&	\textit{\textbf{+53}}	&	+24	&	+24	&	2\\
90569	&	A0 Sr Cr Si	&	\textit{\textbf{+26}}	&	\textit{\textbf{+42}}	&\textit{\textbf{	+57}}	&	+36	&	+26	&	2\\
112185	&	A1 Cr Eu Mn	& +18	&	+35	&	\textit{\textbf{+34}}	& $-$ & +6 & 2\\
112413	&	A0 Eu Si Cr	&	+13	&	+19	&	+22	&	+40	&	+26	&	2\\
118022	&	A2 Cr Eu Sr	&	\textit{\textbf{+43}}	&	\textit{\textbf{+78}}	&	\textit{\textbf{+77}}	&	+51	&	+34	&	2\\
120198	&	A0 Eu Cr Sr	& +1	&	\textit{\textbf{+40}}	&	\textit{\textbf{+39}}	&	+38	&	+31	&	2\\
137909	&	A9 Sr Eu Cr	&	\textit{\textbf{+39}}	&	+21	&	\textit{\textbf{+47}}	&	+25	&	+4	&	2\\
148112	&	A0 Cr Eu	&	$-$6	&	+7	&	+8	&	$-$	&	+13	&	2\\
170000	&	A0 Si	&	+7	&	+6	&	+6	&	$-$	&	$-$	&	2\\
183056	&	B9 Si	&	+13	&	+23	&	+20	&	$-$	&	+2	&	2\\
201601	&	A9 Sr Eu	&	$-$14	&	\textit{\textbf{$-$54}}	&	\textit{\textbf{$-$48}}	&	+10	&	+4	&	2\\
206088	&	A7$-$F3 Sr	&	+1	&	\textit{\textbf{$-$50}}	&	\textit{\textbf{$-$48}}	&	+7	&	+2	&	2\\
\hline
358	&	B9 Mn Hg	&	0	&	$-$7	&	$-$6	&	$-$	&	+5	&	3\\
23950	&	B9 Mn Hg Si	&	+3	&	$-$19	&	$-$14	&	$-$	&	0	&	3\\
33904	&	B9 Hg Mn	&	$-$1	&	+3	&	+3	&	$-$	&	$-$4	&	3\\
35497	&	B8 Cr Mn	&	+3	&	+4	&	+3	&	$-$	&	$-$7	&	3\\
77350	&	B9 Sr Cr Hg	&	$-$4	&	+10	&	+9	&	+1	&	+5	&	3\\
78316	&	B8 Mn Hg	&	0	&	+18	&	+20	&	+12	&	+6	&	3\\
106625	&	B8 Hg Mn &+3	&	$-$18	&	$-$11	&	$-$	&	$-$9	&	3\\
129174	&	B9 Mn Hg  &	$-$4	&	+4	&	+1	&	$-$	&	0	&	3\\
143807	&	A0 Mn Hg  &	$-$7	&	+13	&	+3	&	$-$	&	+6	&	3\\
145389	&	B9 Mn Hg  & +15	&	+8	&	+19	&	$-$	&	+4	&	3\\
220933	&	A0 Hg	Mn &	$-$3	&	$-$2	&	+7	&	$-$	&	$-$5	&	3\\
\hline
11415	&	B3 He wk.	& +13	&	\textit{\textbf{$-$41}}	&	\textit{\textbf{$-$34}}	&	$-$	&	+28	&	4\\
23408	&	B7 He wk. Mn	&	+13	&	$-$9	&	$-$9	&	+5	&	+3	&	4\\
115735	&	B9 He wk.	&	$-$11	&	\textit{\textbf{$-$42}}	&	\textit{\textbf{$-$49}}	&	+2	&	$-$2	&	4\\
\hline 
\end{tabular}
\end{center}
\end{table*}

\begin{table*}
\caption{Synthetic ($\Delta a$, $\Delta a'$, and $\Delta a' mod$) and observed ($\Delta a\,obs$ and $\Delta (V1-G)$)
peculiarity indices in mmags for apparent non-CP stars detected via synthetic photometry. The synthetic 
photometric values for detected objects are given in boldface italics. The upper panel lists well-known
emission-type and metal-weak objects.}
\label{deltaa_values_other}
\begin{center}
\begin{tabular}{lcccccccc} 
\hline\hline 
\\
HD	&	SpType	&	$\Delta a$	&	$\Delta a'$	&	$\Delta a' mod$	&	$\Delta a\,obs$	& $\Delta (V1-G)$\\
\hline
5394	&	B0.5 IVe	&	+17	&	\textit{\textbf{$-$37}}	&	$-$21	& +5 & +1 \\
6811	&	B7 Ve	&	$-$10	&	\textit{\textbf{$-$45}}	&	\textit{\textbf{$-$41}}	&	$-$5 & +2\\
18552	&	B8 Ve	&	\textit{\textbf{+40}}	&	\textit{\textbf{+100}}	&\textit{\textbf{+97}}	&	$-$ & $-$2\\
22192	&	B5 Ve	&	$-$9	&	\textit{\textbf{$-$49}}	&	\textit{\textbf{$-$44}}	&	$-$ & $-$4\\
23016	&	B8V (e)	&	$-$16	&	\textit{\textbf{$-$38}}	&	$-$30	& $-$ & $-$6 \\
32537	&	F1 Vp MgII 4481\AA\,weak	&	$-$10	&	\textit{\textbf{$-$51}}	&	$-$31	& $-$ & $-$7 \\
34078	&	O9.5 Ve, var.	&	+4	&	\textit{\textbf{$-$57}}	&	\textit{\textbf{$-$52}}	&	$-$ & $-$10\\
35439	&	B1 Vpe	&	$-$11	&	$-$34	&	\textit{\textbf{$-$49}}	& +5 &	+3\\
67934	&	A0 Vnp MgII 4481\AA\,weak	&	$-$1	&	\textit{\textbf{+40}}	&	+13	& $-$ & $-$4 \\
74873	&	kA0.5hA5mA0.5V $\lambda$ Boo	&	$-$18	&	\textit{\textbf{$-$37}}	&	$-$25	& $-$ & $-$15 \\
111604	&	A5 Vp $\lambda$ Boo	&	$-$11	&	$-$34	&\textit{\textbf{$-$39}}	&	$-$ & $-$2\\
112014	&	A0 IIsp Mg,Si weak	&	$-$9	&	\textit{\textbf{$-$39}}	&	$-$30	& $-$ & +2 \\
193237	&	B1 ep	&	+6	&	\textit{\textbf{$-$97}}	&	$-$14	& +25 & +17 \\
209409	&	B7 Ive	&	$-$18	&	\textit{\textbf{+53}}	&	+24	&	$-$ & +4\\
210839	&	O6 If(n)p(e)	& +6	&	\textit{\textbf{$-$69}}	&	\textit{\textbf{$-$55}}	&	$-$ & $-$19\\
217891	&	B6 IIIe	&	+4	&	\textit{\textbf{+65}}	&	\textit{\textbf{+48}}	&	$-$1 & +4\\
\hline
17769	&	B7 V	&	+18	&	+41	&	\textit{\textbf{+49}}	&	$-$ & $-$3\\
19374	&	B1 V $\beta$ Cep	&	\textit{\textbf{+50}}	&	\textit{\textbf{+62}}	&	\textit{\textbf{+91}}	&	$-$ & $-$5\\
22951	&	A1 Vn	&	\textit{\textbf{+32}}	&	+33	&	\textit{\textbf{+33}}	& $-$ & $-$7 \\
24554	&	A1 V	& \textit{\textbf{+32}}	&	$-$6	&	$-$10	& $-$ & $-$ & \\
28052	&	F0 V	&	\textit{\textbf{+32}}	&	+17	&	+19	&	$-$ & +6\\
28149	&	B5 V	&	\textit{\textbf{+42}}	&	+28	& $-$ & $-$ \\
29248	&	B2 III $\beta$ Cep & +4	&	\textit{\textbf{+44}}	&	+32	& $-$ & $-$9 \\
35671	&	B5 V SB	&	+18	&	\textit{\textbf{+51}}	&	\textit{\textbf{+35}}	&	$-$ & $-$5\\
35770	&	B9.5 Vn	&	\textit{\textbf{+26}}	&	\textit{\textbf{+40}}	&	\textit{\textbf{+33}}	& $-$ & $-$4 \\
70011	&	B9.5 V	&	$-$2	&	\textit{\textbf{+40}}	&	+22	& $-$ & +2 \\
76582	&	A7 V	&	\textit{\textbf{+24}}	&	+18	&	+25	& $-$ & $-$6 \\
83808	&	F8-G0III + A7m	&	+18	&	+18	&	\textit{\textbf{+34}}	& $-$ & $-$10 \\
107700	&	G7III + A3IV	&	\textit{\textbf{+29}}	&	+1	&	+18	& $-$ & $-$4 \\
119765	&	A0 V	&	$-$7	&	\textit{\textbf{$-$50}}	&	\textit{\textbf{$-$33}}	& +1 & $-$ \\
139891	&	B6V SB2	&		+11	&	\textit{\textbf{+41}}	&	+28	& $-$ & $-$ \\
166182	&	B2 IV	&	+6	&	\textit{\textbf{+38}}	&	+27	& $-$ & $-$3 \\
178596	&	F2 IV-V	&	$-$3	&	\textit{\textbf{$-$45}}	&	$-$28	& $-$ & $-$8 \\
188260	&	B9.5 III	&	$-$15	&	\textit{\textbf{$-$47}}	&	\textit{\textbf{$-$38}}	&	$-$ & $-$\\
188350	&	A0 III	&	$-$13	&	\textit{\textbf{$-$49}}	&	\textit{\textbf{$-$42}}	&	$-$ & $-$8\\
212120	&	B6 V, ell. var.	&	\textit{\textbf{$-$45}}	&	$-$15	&	\textit{\textbf{$-$41}}	&	$-$ & $-$2\\
222603	&	A7 V	&	+7	&	\textit{\textbf{$-$69}}	&	\textit{\textbf{$-$36}}	&	$-$5 & $-$4\\
\hline 
\end{tabular}
\end{center}
\end{table*}

\subsection{Apparent normal-type objects} \label{normal_sub}

Table \ref{deltaa_values_other} lists the apparently non-CP stars detected via synthetic photometry.  
This sample
can be divided into emission-type and metal-weak objects as well as inconspicuous stars.

The hot-emission-type stars are defined as dwarfs that have shown
hydrogen emission in their spectra at least once. Due to a developed
equatorial disk that is produced by stellar winds, emission arises
quite regularly. In addition, photometric variability on different
time scales is a common phenomenon caused by the formation
of shock waves within those disks. The phases of emission are replaced by shell and normal
phases of the same object, leading to a transition from negative to positive $\Delta a$ values
\citep{Pavl89}. Among the metal-weak objects, the most
prominent subgroup are the $\lambda$ Bootis stars. This small group of objects comprises late-B 
to early-F-type stars, with moderate to extreme (up to a factor 100) surface under-abundances of 
most Fe-peak elements and solar abundances of lighter elements (C, N, O, and S). The main mechanisms 
responsible for this phenomenon are atmospheric diffusion, meridional mixing, and accretion of material 
from their surroundings \citep{Paun02}. The $\Delta a$ observations of the two groups are summarized 
in \citet{Paun05}.

In the sample of inconspicuous objects, we detected two $\beta$ Cephei pulsators (HD 19374 and HD 29248) 
with significantly high 
positive $\Delta a$ values. These objects are early-type-B stars with light and
radial velocity variations on time scales of several hours \citep{stak05}. Since these objects are quite
rare, fewer than 200 Galactic objects are known, it would be interesting to find out whether they can be detected via  
$\Delta a$ photometry. Unfortunately, no observations in this respect have been performed so far.

The remaining objects have two characteristics in common:
\begin{enumerate}
\item Except for one object (HD 22951), all of them are very fast rotators ($v \sin i$\,$\geq$\,150\,km s$^{-1}$)
which is atypical for CP stars.
\item Almost all objects are in binary systems, which might distort the spectrophotometry.
\end{enumerate}

For the following objects we found additional interesting characteristics in the literature:
\begin{itemize}
\item {\it HD 24554}: \citet{schr07} found strong X-ray emission indicating an undetected binary nature
of the object.
\item {\it HD 28052}: this is the second-brightest X-ray source in the Hyades and a possibly quadruple system
with at least one component of $\delta$ Scuti type \citep{simo00}.
\item {\it HD 83808}: this spectroscopic binary system is listed as CP1 candidate in \citet{rens09}. 
\item {\it HD 107700}: \citet{Grif11} analysed this close-binary system and found a slight metal-weakness
for both components.
\item {\it HD 119765}: \citet{rens09} listed it as questionable CP candidate without any designation to
a specific subgroup.
\end{itemize}
For these samples, $\Delta a'$ is superior to the other two systems. 
For almost all of the newly discovered peculiar objects among the normal-type objects, photometric 
$\Delta a$ and/or spectroscopic observations are needed to clarify their nature.

We also identified nine cool-type objects (0.5\,$<$\,$(B-V)_0$\,$<$\,1.5\,mag) that significantly
deviate in the positive direction. Those stars are
\begin{itemize}
\item {\it HD 19373}: \citet{Cant11} analysed the chromospheric activity of the G0\,V object.
\item {\it HD 26965}: this is a young triple binary system including flare-type objects \citep{Pett91}.
\item {\it HD 35369}: \citet{Prug11} found an underabundance of [Fe/H]\,=\,$-$0.22\,dex compared with the Sun.
\item {\it HD 49878}: there are no detailed investigations for this K-type giant available in the literature.
\item {\it HD 68375}: \citet{take08} investigated it in more detail, and found no peculiarities. 
\item {\it HD 82635}: this is a highly chromospherically active RS CVn type giant \citep{stra94}. 
\item {\it HD 158899}: according to \citet{Anti04}, this is a moderate Barium star classified as K3.5\,III\,Ba0.1.
This group of chemically peculiar stars consists of G to K giants, whose spectra indicate an overabundance of 
s-process elements. 
\item {\it HD 192577}: it is a $\zeta$ Aurigae type eclipsing binary of spectral type K4\,I \citep{Eato08}.
\item {\it HD 194093}: \citet{Gray10} presented photometric and spectroscopic time series of this variable star. 
The variations are found on all time scales up to several hundred days.
\end{itemize}
We found that the $a$ indices are linearly correlated with the effective temperature up to a spectral type of
M0. For cooler-type objects, there is a strong indication of an additional luminosity effect that is superimposed on the
temperature dependency.
Up to now, no photometric $\Delta a$ data of cool-type stars have been published. Our findings suggest
that such observations will be very interesting for the study of low-mass peculiar objects.

\section{Conclusions and outlook}

We used the spectrophotometric data of the stellar catalogue by \citet{Khar88} to synthesize three different
``$a$ systems''. These data cover the complete spectral range from low- to high-mass objects. For the first time, we
presented  $a$ indices for stars cooler than F-type. Excluding low-quality data and early-type supergiants,
1067 stars were used to synthesize different peculiarity indices. Our main results are
\begin{itemize}
\item Most of the known classical magnetic chemically peculiar stars were detected.
\item We presented a list of about 50 normal-type objects across the complete spectral range that
were detected.
\item The most efficient $a$ system is very similar to that previously employed by \citet{Mait80a}.
\item The normality line of the $a$ system correlates with the effective temperature up to a spectral type of M0.
\end{itemize}
Our analysis showed that spectrophotometric data can be used for calculating synthetic $a$ indices and
for detecting peculiar objects across the complete spectral range up to M0. As next steps, we will observe cool-type 
objects to establish the corresponding correlations. In addition, we will use classification
resolution spectra (for example from LAMOST) to search for chemically peculiar objects in a semi-automatic
way.

\begin{acknowledgements}
This project is financed by the SoMoPro II programme (3SGA5916). The research leading
to these results has acquired a financial grant from the People Programme
(Marie Curie action) of the Seventh Framework Programme of EU according to the REA Grant
Agreement No. 291782. The research is further co-financed by the South-Moravian Region. 
It was also supported by the grants GA \v{C}R P209/12/0217, 14-26115P, 7AMB12AT003, and
the financial contributions of the Austrian Agency for International 
Cooperation in Education and Research (BG-03/2013 and CZ-10/2012).
This research has made use of the WEBDA database, operated at the Department of 
Theoretical Physics and Astrophysics of the Masaryk University.
This work reflects only the author's views and the European 
Union is not liable for any use that may be made of the information contained therein.
\end{acknowledgements}


\begin{thebibliography}{}
\bibitem[\protect\citeauthoryear{Adelman}{1979}]{Adel79} Adelman, S. J. 1979, \aj, 84, 857
\bibitem[\protect\citeauthoryear{Adelman et al.}{1989}]{Adel89} Adelman, S. J., Pyper, D. M., Shore, S. N., White, R. E., Warren, W. H., Jr.
1989, \aaps, 81, 221 
\bibitem[\protect\citeauthoryear{Adelman et al.}{1999}]{Adel99} Adelman, S. J., Caliskan, H., Cay, T., Kocer, D., \&
Tektanali, H. G. 1999, \mnras, 305, 591
\bibitem[\protect\citeauthoryear{Antipova et al.}{2004}]{Anti04} Antipova, L. I., Boyarchuk, A. A., Pakhomov, Yu. V., Panchuk, V. E.
2004, Astronomy Reports, 48, 597
\bibitem[\protect\citeauthoryear{Babcock}{1947}]{Babc47} Babcock, H. W. 1947, \apj, 105, 105
\bibitem[\protect\citeauthoryear{Burnashev}{1985}]{Burn85} Burnashev, V. I. 1985, Abastumanskaya Astrofiz. Obs., Byull., 59, 83
\bibitem[\protect\citeauthoryear{Bychkov et al.}{2009}]{Bych09} Bychkov, V. D., Bychkova, L. V., \& Madej, J. 2009, \mnras, 394, 1338
\bibitem[\protect\citeauthoryear{Canto Martins et al.}{2011}]{Cant11} Canto Martins, B. L., das Chagas, M. L., Alves, S. et al.,
2011, \aap, 530, A73
\bibitem[\protect\citeauthoryear{Catalano et al.}{1998}]{Cata98} Catalano, F. A., Leone, F., \& Kroll, R. 1998, \aaps, 129, 463
\bibitem[\protect\citeauthoryear{Cramer}{1999}]{Cram99} Cramer, N. 1999, \na, 43, 343
\bibitem[\protect\citeauthoryear{Eaton}{2008}]{Eato08} Eaton, J. A. 2008, \aj, 136, 1964 
\bibitem[\protect\citeauthoryear{Glagolevskij}{2013}]{Glag13} Glagolevskij, Yu. V. 2013, Astrophysics, 56, 173
\bibitem[\protect\citeauthoryear{Glushneva et al.}{1992}]{Glus92} Glushneva, I. N., Kharitonov, A. V., Kniazeva, L. N., \& Shenavrin, V. I.
1992, \aaps, 92, 1
\bibitem[\protect\citeauthoryear{Gray}{2010}]{Gray10} Gray, D. F. 2010, \aj, 140, 1329
\bibitem[\protect\citeauthoryear{Griffin \& Griffin}{2011}]{Grif11} Griffin, R. E. M., \& Griffin, R. F. 2011, Astron. Nachr., 332, 105 
\bibitem[\protect\citeauthoryear{Gutierrez-Moreno}{1975}]{Guti75} Gutierrez-Moreno, A. 1975, \pasp, 87, 805
\bibitem[\protect\citeauthoryear{Hauck}{1974}]{Hauc74} Hauck, B. 1974, \aap, 32, 447
\bibitem[\protect\citeauthoryear{Havnes \& Conti}{1971}]{Havn71} Havnes, O., \& Conti, P. S. 1971, \aap, 14, 1
\bibitem[\protect\citeauthoryear{Iliev et al.}{2006}]{Ilie06} Iliev, I. Kh., Budaj, J., Fenovc{\' i}k, M., Stateva, I., \& Richards, M. T.
2006, \mnras, 370, 819
\bibitem[\protect\citeauthoryear{Jordi et al.}{2010}]{Jord10} Jordi, C., Carrasco, J. M., Fabricius, C., Figueras, F., \& Voss, H. 2010,
Highlights of Spanish Astrophysics V, Astrophysics and Space Science Proceedings, Springer-Verlag, Berlin, p. 147
\bibitem[\protect\citeauthoryear{Khan \& Shulyak}{2007}]{Khan07} Khan, S. A., \& Shulyak, D. V. 2007, \aap, 469, 1083
\bibitem[\protect\citeauthoryear{Kharitonov et al.}{1988}]{Khar88} Kharitonov, A. V., Tereshchenko, V. M., \& 
Knyazeva, L. N. 1988, The spectrophotometric catalogue of stars, Nauka, Alma-Ata (USSR)
\bibitem[\protect\citeauthoryear{Kodaira}{1969}]{Koda69} Kodaira, K. 1969, \apj, 157, L59
\bibitem[\protect\citeauthoryear{Kohoutek \& Wehmeyer}{1999}]{Koho99} Kohoutek, L., \& Wehmeyer, R. 1999, \aaps, 134, 255
\bibitem[\protect\citeauthoryear{Maitzen}{1976}]{Mait76} Maitzen, H. M. 1976, \aap, 51, 223
\bibitem[\protect\citeauthoryear{Maitzen}{1980}]{Mait80a} Maitzen, H. M. 1980, \aap, 89, 230
\bibitem[\protect\citeauthoryear{Maitzen \& Moffat}{1972}]{Mait72} Maitzen, H. M., \& Moffat, A. F. J. 1972, \aap, 16, 385
\bibitem[\protect\citeauthoryear{Maitzen \& Muthsam}{1980}]{Mait80b} Maitzen, H. M., \& Muthsam, H. 1980, \aap, 83, 334
\bibitem[\protect\citeauthoryear{Mermilliod et al.}{1997}]{Merm97} Mermilliod, J.-C., Mermilliod, M., \& Hauck, B. 1997, \aaps, 124, 349 
\bibitem[\protect\citeauthoryear{Napiwotzki et al.}{1993}]{na93} Napiwotzki, R., Schoenberner, D., \& Wenske, V. 1993, \aap, 268, 653 
\bibitem[\protect\citeauthoryear{Netopil et al.}{2008}]{net08} Netopil, M., Paunzen, E., Maitzen, H. M., North, P., \& Hubrig, S. 2008, \aap, 491, 545 
\bibitem[\protect\citeauthoryear{Paunzen et al.}{2002}]{Paun02} Paunzen, E., Iliev, I. Kh., Kamp, I., \& Barzova, I. 2002, \mnras, 336, 1030
\bibitem[\protect\citeauthoryear{Paunzen et al.}{2005}]{Paun05} Paunzen, E., St{\"u}tz, Ch., \& Maitzen, H. M. 2005, \aap, 441, 631
\bibitem[\protect\citeauthoryear{Pavlovski \& Maitzen}{1989}]{Pavl89} Pavlovski, K., \& Maitzen, H. M. 1989, \aaps, 77, 351
\bibitem[\protect\citeauthoryear{Perraut et al.}{2011}]{Perr11} Perraut, K., Brand{\~a}o, I., Mourard, D. et al., 2011, \aap, 526, A89
\bibitem[\protect\citeauthoryear{Pettersen}{1991}]{Pett91} Pettersen, B. R. 1991, \memsai, 62, 217
\bibitem[\protect\citeauthoryear{Preston}{1974}]{Pres74} Preston, G. W. 1974, \araa, 12, 257
\bibitem[\protect\citeauthoryear{Prugniel et al.}{2011}]{Prug11} Prugniel, P., Vauglin, I., \& Koleva, M. 2011, \aap, 531, A165
\bibitem[\protect\citeauthoryear{Renson \& Manfroid}{2009}]{rens09} Renson, P., \& Manfroid, J. 2009, \aap, 498, 961
\bibitem[\protect\citeauthoryear{Schr{\"o}der \& Schmitt}{2007}]{schr07} Schr{\"o}der, C., \& Schmitt, J. H. M. M. 2007, \aap, 475, 677
\bibitem[\protect\citeauthoryear{Simon \& Ayres}{2000}]{simo00} Simon, T. \& Ayres, T. R. 2000, \apj, 539, 325
\bibitem[\protect\citeauthoryear{Sokolov}{2006}]{Soko06} Sokolov, N. A. 2006, \mnras, 373, 666
\bibitem[\protect\citeauthoryear{Stankov \& Handler}{2005}]{stak05} Stankov, A., \& Handler, G. 2005, \apjs, 158, 193
\bibitem[\protect\citeauthoryear{Stift \& Alecian}{2012}]{Stif12} Stift, M. J., \& Alecian, G. 2012, \mnras, 425, 2715
\bibitem[\protect\citeauthoryear{Strassmeier et al.}{1994}]{stra94} Strassmeier, K. G., Handler, G., Paunzen, E., Rauth, M. 1994, \aap, 281, 855
\bibitem[\protect\citeauthoryear{Takeda et al.}{2008}]{take08} Takeda, Y., Sato, B., \& Murata, D. 2008, \pasj, 60, 781
\end{thebibliography}
\end{document}